\documentclass[aps,prd,reprint,onecolumn,notitlepage,superscriptaddress]{revtex4-1}

\usepackage{lipsum}
\usepackage{amsmath}
\usepackage{graphicx}
\usepackage{color}
\usepackage{tikz}
\usepackage{verbatim}

\usepackage{amssymb}
\usepackage{bm}

\newcommand{\rom}[1]{\textup{\uppercase\expandafter{\romannumeral#1}}}

\begin{document}

\title{Higher Derivative Theory For Curvature Term Coupling With Scalar Field}
\author{Pawan Joshi}
\email{pawanjoshi697@iiserb.ac.in}
\author{ Sukanta Panda}
\email{sukanta@iiserb.ac.in}
\affiliation{Department of Physics, Indian Institute of Science Education and Research, Bhopal, India}

%\date{\today}

\begin{abstract}
Higher order derivative theories, generally suffer from  instabilities, known as
Ostrogradsky instabilities. This issue can be resolved by removing any existing degeneracy present in
such theories. We consider a model involving at most second order derivatives of scalar field
non-minimally coupled to curvature terms. Here we perform (3+1) decomposition of Lagrangian to separate second order time derivative form the rest. This is useful to check the degeneracy hidden in the Lagrangian will help us to find conditions under which Ostrogradsky instability do not appear. In our case, we find no such non trivial conditions which can stop the appearance of the Ostrogradsky ghost.
\end{abstract}

\maketitle
\section{INTRODUCTION}
Observations suggest that our current universe is in an accelerating phase\cite{RiessAdamG:1998cb} and explanation of this acceleration can be provided by dark energy. The dark energy is generally modeled by modifying gravity part of Einstein Hilbert action\cite{Nojiri:2006ri}. One of the brands of this modified gravity model is scalar-tensor theories\cite{Gleyzes:2014rba, PhysRevD.37.3406}. 
When we consider scalar-tensor theories with higher order derivative terms. If higher derivative  Lagrangian is non-degenerate there exist a ghost-like instability known as Ostrogradsky 
instability\cite{Ostrogradsky:1850fid} in which  Hamiltonian contains such terms which are linear in momentum. 
In degenerate theory, the higher derivative is present in the Lagrangian but they cancel such a way that does not appear in the equation of motion\cite{Nicolis:2008in}.
In 1974 Horndeski proposed a general action for scalar field that contain the higher derivative term in the Lagrangian but the equation of motion is of second order\cite{Horndeski1974}. Horndeski Lagrangian takes a form,
\begin{eqnarray}
 L=  L^{H}_{2}+ L^{H}_{3}+ L^{H}_{4}+ L^{H}_{5},\end{eqnarray}
 where,
\begin{eqnarray}
  L^{H}_{2}=G_{2}(\phi ,X), \qquad
    L^{H}_{3}=G_{3}(\phi ,X) \Box{\phi}, \qquad
     L_{4}^{H}=G_{4}(\phi,X) R - 2G_{4,X}(\phi, X)(   \Box\phi^{2} - \phi_{\mu\nu} \phi^{\mu\nu}),\label{a1} \\  L_{5}^{H}=G_{5}(\phi,X) G_{\mu\nu}\phi^{\mu\nu}+\frac{1}{3}G_{5}(\phi,X) R - (   \Box\phi^{3} - 3  \Box{\phi} \phi_{\mu\nu} \phi^{\mu\nu}+ 2 \phi_{\mu\nu} \phi_{\rho}^{\nu} \phi^{\mu \rho}), \label{a2}  \end{eqnarray}
and $ \phi_{\mu}= \nabla_{\mu} \phi $,  $ \phi_{\mu\nu}= \nabla_{\mu} \nabla_{\nu} \phi $, $ X=\nabla_{\mu} \phi  \nabla^{\mu} \phi $, R, $R_{\mu\nu}$, $G_{\mu\nu}$ is Ricci scalar, Ricci Tensor and Einstein tensor respectively.  Here we notice that In $L_{2}^{H}$ is some combination scalar field and its first derivative and $L_{3}^{H}$ additionally contain the second derivative of the scalar field, $L_{4}^{H}$, $L_{5}^{H}$ contain curvature term and first and second derivative of $ \phi$. In the case, we have constructed a Lagrangian containing higher derivatives of a scalar field with non-minimal coupling to curvature. Our motivation is to find a degeneracy condition to get rid of Ostrogradsky instability in our higher derivative Lagrangian.
\section{POSSIBLE TERMS FOR $\nabla_{\mu}\nabla_{\nu} \phi \nabla_{\rho}\nabla_{\sigma} \phi $}
Consider an action of the form,
\begin{eqnarray}
 S= \int d^{4}x \sqrt{-g} \tilde{C}^{\mu \nu ,\rho \sigma}  \nabla_{\mu} \nabla_{\nu} \phi \nabla_{\rho} \nabla_{\sigma} \phi,
\end{eqnarray} 
where $\tilde{C}^{\mu \nu ,\rho \sigma} $ contain metric tensor and curvature term and it is  only possible term for non-minimal coupling of second derivative of scalar field with the curvature term. Its simple form is,
\begin{eqnarray}
\tilde{C}^{\mu \nu ,\rho \sigma}= (D_{1} g^{\mu \rho} g^{\nu \sigma} +D_{2} g^{\mu \sigma} g^{\nu \rho}) R  + (D_{3} g^{ \eta \rho} g^{ \mu \nu } g^{\beta \sigma}  +D_{4}  g^{\mu \eta} g^{\beta \rho} g^{\nu \sigma})R_{\eta \beta} \nonumber  + (D_{5}  g^{ \mu \eta } g^{  \nu \beta } g^{\gamma \rho} g^{ \delta \sigma} + D_{6} g^{  \mu \eta} g^{ \sigma \beta } g^{\gamma \rho} g^{\delta \nu}) R_{\eta \beta \gamma \delta}.
\end{eqnarray}
 Now we rewrite the action replace first derivative of scalar field by a new field $A_{\mu}$ i.e. $\nabla_{\mu} \phi=A_{\mu} $. So the new form of action is given as,
\begin{eqnarray}
 S= \int d^{4}x \sqrt{-g} [\tilde{C}^{\mu \nu ,\rho \sigma}  \nabla_{\mu} A_{\nu}  \nabla_{\rho} A_{\sigma} +\lambda^{\mu} (\nabla_{\mu} \phi-A_{\mu})],
\end{eqnarray}
this action can be written as,
\begin{eqnarray}
 S= \int d^{4}x \sqrt{-g} [\mathcal{L}_1+\mathcal{L}_2+\mathcal{L}_3+\mathcal{L}_4+\mathcal{L}_5+\mathcal{L}_6]+\lambda^{\mu} (\nabla_{\mu} \phi-A_{\mu})],\label{1s}
\end{eqnarray} where,
\begin{eqnarray}
\mathcal{L}_{1} &=& D_{1} R  g^{\mu \rho} g^{\nu \sigma}  \nabla_{\mu} A_{\nu} \nabla_{\rho} A_{\sigma}, \\
\mathcal{L}_{2} &=& D_{2}R g^{\mu \nu} g^{\rho \sigma}  \nabla_{\mu} A_{\nu}  \nabla_{\rho} A_{\sigma}, \\
\mathcal{L}_{3} &=& D_{3} g^{ \eta \mu} g^{\beta \rho} g^{\nu \sigma} R_{ \eta \beta}  \nabla_{\mu} A_{\nu} \nabla_{\rho} A_{\sigma}, \\ 
\mathcal{L}_{4} &=& D_{4} g^{ \eta \rho} g^{\mu \nu} g^{\beta \sigma} R_{ \eta \beta} \nabla_{\mu} A_{\nu}   \nabla_{\rho} A_{\sigma}, \\
\mathcal{L}_{5} &=& D_{5}g^{\mu  \eta} g^{\nu \beta} g^{\gamma \rho} g^{\delta \sigma} R_{ \eta \beta \gamma \delta} \nabla_{\mu} A_{\nu} \nabla_{\rho} A_{\sigma}, \\
\mathcal{L}_{6} &=& D_{6}g^{\mu  \eta} g^{\nu \beta} g^{\gamma \rho} g^{\delta \sigma} R_{ \eta \beta \gamma \delta}\nabla_{\mu} A_{\nu} \nabla_{\rho} A_{\sigma},
\end{eqnarray}
and $\lambda^{\mu}$ are Lagrange multiplier. Here it is noted that due to the symmetry property of Riemann tensor $L_5$ term vanishes.
\section{Representation of action in (3+1) decomposition} 
In this section, we want to find degeneracy condition on Lagrangian, using a (3+1) decomposition.  Our (3+1) convention and  notation similar to ref.\cite{baumgarte2010numerical,langlois2016degenerate}. 
 in (3+1) decomposition  $\nabla_{a} A_{b} $ is,  
\begin{eqnarray}
\nabla_{a} A_{b} &=& \mathcal{D}_{a} \mathcal{A}_{b} - A_{*} K_{a b} +n_{a}(K_{b c} \mathcal{A}^{c} -\mathcal{D}_{b} A_{*} )+ n_{b}(K_{a c} \mathcal{A}^{c} -\mathcal{D}_{a} A_{*} )+  \frac{1}{N} n_{a} n_{b}(\dot{A}_{*} -N^{c} D_{c} A_{*}-N \mathcal{A}_{c} a^{c}).
\end{eqnarray}
Now we introduced $ X_{a b}=(D_{a} \mathcal{A}_{b} - A_{*} K_{a b} $), $ Y_{b}=(K_{b c} \mathcal{A}^{c} -D_{b} A_{*}$) and $ Z=(N^{c} D_{c} A_{*}+N \mathcal{A}_{c} a^{c}$) for simplifying  the calculation so eq.(13) becomes,
\begin{eqnarray}
 \nabla_{a} A_{b}&=& \dfrac{1}{N}  n_{a} n_{b} (\dot{A}_{*}-Z)-X_{ab}-n_{a}Y_{b} \ -n_{b} Y_{a},
\end{eqnarray}
where $ \mathcal{D}_a $ spatial derivative associated with spatial metric $ h_{ab}$. $ \mathcal{A}_a $ and $A _{*}$  spatial  and normal projection of  4-vector $A_{a}$ given as $ \mathcal{A}_{a}=h_{a}^{b}A_{b}  $ and $A_{*}=n^{a}A_{a}$, $N^{a}$ is shift vector, N is lapse function,
$a_b = n^{c}\nabla_{c}n_b $ is the acceleration vector and $ K_{a b}$ is Extrinsic curvature tensor related to first derivative of metric. In (3+1) formalism $  R,R_{\mu \nu},R_{\mu\nu\rho\sigma} $ are, 
\begin{eqnarray}
R &=& \mathcal{R}+ K^{2}-3 K_{a b} K^{a b}+ 2 h^{a b} L_{n}  K_{a b} -2 \mathcal{D}_{b} a^{b}-2a_{b} a^{b},
 \\{}_{\perp}R_{ab} &=& \mathcal{R}_{a b}+ K_{a b} K - K_{a s} K^{s}_{b}+  L_{n}  K_{a b} - \mathcal{D}_{a} a_{b}- a_{a} a_{b},\\
{}_{\perp}R_{b n} &=& \mathcal{D}_{s}K_{b}^{s}-\mathcal{D}_{b} K,  \\
{}_{\perp}R_{n n} &=&  K_{s t} K^{s t}- h^{s t} L_{n}  K_{s t} + \mathcal{D}_{s} a^{s}+a_{s} a^{s},\\
{}_{\perp} R_{a b c d} &=& \mathcal{R}_{a b c d}+K_{a c} K_{a d}- K_{a d}K_{b c}, \\
{}_{\perp} R_{a b c n}  &=& \mathcal{D}_{a}K_{b c}-\mathcal{D}_{b}K_{a c}, \\
{}_{\perp} R_{a b n n} &=&  K_{a u} K^{u}_{b}- L_{n}  K_{a b} + \mathcal{D}_{a} a_{b}-a_{a} a_{b}, \end{eqnarray}
where $ L_{n}K_{a b}$ is Lie derivative of  Extrinsic curvature tensor and related to second order derivative of metric. $ {}_{\perp}R_{ab},{}_{\perp}R_{b n},{}_{\perp}R_{n n} $ is spatial, one normal and two normal projection of Ricci tensor and $ {}_{\perp} R_{a b c d},{}_{\perp} R_{a b c n},{}_{\perp} R_{a b n n} $ is spatial, one normal and two normal projection of Riemann tensor known as Gauss, Codazzi and  Ricci relations respectively.\\
 By using these relation we decompose the action (\ref{1s}) in (3+1) formalism and  separate out second order derivative of metric. Next we derive conditions that no second order derivative of metric appear in the action. Here we are not analysing Ostrogradsky instability arising from the higher derivative of scalar field.
\section{CONDITION FOR NO SECOND ORDER DERIVATIVE OF METRIC}In this section, after substituting (13-20) in eq.(\ref{1s}) and keeping the terms that are second derivative of metric. 
\begin{eqnarray} \mathcal{L}= h^{a b} L_{n}  K_{a b}( \frac{\dot{A}_{*}^2}{N^2}-\frac{2 \dot{A}_{*} Z}{N^2}+\frac{Z^2}{N^2})(2D_{1}+2D_{2}+D_{3}+D_{4})+h^{a b}L_{n}K_{a b}Y_{c}Y^{c}(-4D_1-D_3)+h^{ab}L_{n}K_{a b}\nonumber\\\frac{ (\dot{A}_{*}-Z)  X}{N}( 4D_2+D_4)+2h^{a b} L_{n}K_{a b}( D_{1} X^{cd} X_{cd}+D_{2}X^2) +L_{n}K_{a b}Y^{a}Y^{b}(-D_{3}-2D_{6})\nonumber\\+L_{n}K_{a b}\frac{(\dot{A}_{*}-Z)  X^{ab}}{N}( D_4-2D_6) + L_{n}K_{a b}(D_{3} X_{d}^{a} X^{ b d} +D_{4}X X^{a b})+\textnormal{other terms}. 
 \end{eqnarray}
 To be free from terms containing second derivative of metric in the Lagrangian, we require the coefficient of $ h^{a b} L_{n}  K_{a b}$ and $ L_{n}  K_{a b}$ to vanish. This amounts to a trivial condition $D_1=D_2=D_3=D_4=D_5=D_6=0$. There is no non trivial condition exists in this case.

\section{Unitary Gauge} Here we check the possibility to get rid of second derivative of metric in unitary gauge, this gauge gives the condition$ \phi(x,t)=\phi_{o}(t) $. In this case constant time hypersurfaces coincide with uniform scalar field hypersurfaces.
When we apply unitary gauge ($\mathcal{A}^{a}=0 $), demand that the values of $ X_{a b}$, $ Y_{b}$ and Z become,  $ X_{a b}= - A_{*} K_{a b} $, 
$ Y_{b}=-D_{b} A_{*}$
and $ Z=0$ respectively, after substituting this result into eq.(22), we get
\begin{eqnarray}
\mathcal{L} = h^{a b} L_{n}  K_{a b}\frac{\dot{A}_{*}^{2}} {N^{2}}(2D_{1}+2D_{2}+D_{3}+D_{4})+\frac{A_{*}}{N} \dot{A}_{*} K_{c}^{c}( 4D_2+D_4) +2h^{a b} L_{n}  K_{a b}A_{*}^{2}( D_{1} K_{c}^{d} K_{d}^{c}+D_{2}A_{*}^{2} K_{c}^{c}K_{d}^{d})\nonumber \\ +  D_{4} L_{n}  K_{a b}\frac{A_{*}}{N} \dot{A_{*}} K^{a b}( D_4+2D_6)  + L_{n}  K_{a b}(D_3A_{*}^{2} K_{c}^{a} K^{ b c} +D_4A_{*}^{2} K_{c}^{c} K^{a b})+\textnormal{other terms}. 
\end{eqnarray}
In this case also, we have find condition on  the condition $D_1=D_2=D_3=D_4=D_5=D_6=0$. There is no non trivial condition exists in this case.
\section{CONCLUSION}
In this paper, we work with higher derivative model where both second derivative of metric and scalar field arise in the Lagrangian. Then using (3+1) decomposition, we have shown that no non trivial conditions can be found under which all the terms containing second order derivative of metric disappear from the Lagrangian.
\section{ACKNOWLEDGEMENT}
This work was partially funded by DST (Govt. of India), Grant No. SERB/PHY/2017041. Calculations were
performed using xAct packages of Mathematica.
 \bibliography{ref}

\end{document}